\documentclass[12pt,reqno]{amsart}

\usepackage{amsmath,amsfonts,amssymb,amscd,amsthm,amsbsy,xcolor,wasysym}
\usepackage{comment}
\usepackage{wrapfig}
\usepackage{fancybox}
\usepackage{multimedia}
\usepackage{graphicx,epsfig}
\usepackage{epstopdf}
\usepackage{url}
\usepackage{bm}
\usepackage{cite}

\numberwithin{equation}{section}

\newtheorem{meta-thm}[theorem]{Meta-Theorem}

\newcommand\beq[1]{ \begin{equation}\label{#1} }
\newcommand{\eeq}{ \end{equation} }

\newcommand\beqa[1]{ \begin{eqnarray} \label{#1}}

\newcommand{\eeqa}{ \end{eqnarray} }
\newcommand{\beqano}{ \begin{eqnarray*} }
\newcommand{\eeqano}{ \end{eqnarray*} }
\newcommand\equ[1]{{\rm (\ref{#1})}}

\begin{document}


\title[The dynamics of the de Sitter resonance]
{The dynamics of the de Sitter resonance}

\author[A. Celletti]{Alessandra Celletti}

\address{
Department of Mathematics, University of Rome Tor Vergata, Via
della Ricerca Scientifica 1, 00133 Roma (Italy)}

\email{celletti@mat.uniroma2.it}

\author[F. Paita]{Fabrizio Paita}

\address{
Department of Mathematics, University of Rome Tor Vergata, Via
della Ricerca Scientifica 1, 00133 Roma (Italy)}

\email{paita@mat.uniroma2.it}

\author[G. Pucacco]{Giuseppe Pucacco}

\address{
Department of Physics, University of Rome Tor Vergata, Via della
Ricerca Scientifica 1, 00133 Roma (Italy)}

\email{pucacco@roma2.infn.it}


\baselineskip=18pt              


\begin{abstract}

We study the dynamics of the de Sitter resonance, namely the
stable equilibrium configuration of the first three Galilean
satellites. We clarify the relation between this family of
configurations and the more general Laplace resonant states. In
order to describe the dynamics around the de Sitter stable
equilibrium, a one-degree of freedom Hamiltonian normal form is
constructed and exploited to identify initial conditions leading
to the two families.

The normal form Hamiltonian is used to check the accuracy in
the location of the equilibrium positions. Besides, it gives a
measure of how sensitive it is with respect to the different
perturbations acting on the system. By looking at the phase-plane
of the normal form, we can identify a \sl Laplace-like \rm
configuration, which highlights many substantial aspects of the
observed one.
\end{abstract}


\maketitle

\section{Introduction}

The three Galilean satellites of Jupiter, Io, Europa and Ganymede
are phase-locked in the so-called \sl Laplace resonance \rm
\cite{FeBook,MD}. This fascinating dynamical state involves the
commensurability 4:2:1 of the mean motions and a locking of the
relative precession of the \sl peri-Jove \rm of Io and Europa. On
the other hand, the peri-Jove of Ganymede is not locked: hence,
out of the four resonant angles combining longitudes and arguments
of peri-Joves, three are \sl librating \rm and one is \sl rotating
\rm \cite{ShMa}.

However, a state in which all four combination angles are
librating is conceivable and indeed dynamically possible. Its
discovery is usually attributed to de Sitter \cite{Si} and is
actually only one of a possible set of dynamical states
\cite{BH}. In a simplified planar model of the mutual interactions of
the three satellites in the Newtonian field of Jupiter, after reducing to four degrees of freedom, the
de Sitter dynamical states are indeed equilibrium points.
In the four degrees of freedom model, the Laplace state corresponds to a periodic orbit
in which three angles are fixed and a fourth angle changes periodically.
The dynamics become quasi-periodic in the full system with six degrees of freedom.
We refer to \sl the \rm de Sitter state as the
only \sl stable \rm equilibrium in all four angular variables.

The starting model considered by de Sitter in \cite{Si} is the planar 4-body problem Jupiter-Io-Europa-Ganymede, in which the influence
of Callisto and the Sun is neglected. There, he considers periodic
orbits of $1^{st}$ and $2^{nd}$ kind, whose solutions, according
to \cite{Poi}, are circles or ellipses. In \cite{Si} de Sitter proves
the existence of a family of periodic orbits with nearly circular Keplerian
ellipses, parameterized by the eccentricity of one of the satellites. The seminal work of de Sitter was recently re-considered in \cite{ZHB,BH}.
In both works, a 5-body problem including Callisto is considered.
The existence of a positive measure set of Lagrangian invariant tori
is proved to exist in a neighborhood of the periodic orbits.

In the present work we consider the 4-body problem Jupiter-Io-Europa-Ganymede.
This system is described by a Hamiltonian function with 8 degrees of freedom (hereafter, DOF), which
can be reduced to 6 DOF due to the translational symmetry. Following \cite{He,Ma}, we make a
transformation of coordinates, taking into account the  linear combinations of the angles
corresponding to the Laplace resonance. It turns out that two variables are cyclic,
thus reducing the model to a 4-DOF Hamiltonian. We then locate the stable equilibria of this system.
By means of the Lie transform method, we construct a resonant normal form for the Laplace resonance. With this procedure we obtain
an approximation of the Hamiltonian
(namely, an expansion around reference values up to second order in the momenta),
which allows us to reduce the problem to a 1-DOF Hamiltonian. In this context,
we aim at clarifying the interrelationship
between the two dynamical states (de Sitter and Laplace) which can be viewed as explicit
solutions of the 1-DOF Hamiltonian system,
whose normal form has the appearance of a
fundamental resonance problem. On the phase-cylinder of this
Hamiltonian, the de Sitter stable equilibrium is an elliptic fixed
point. The libration domain around it is bounded by a critical curve beyond which
rotating solutions exist (compare with Fig.\ref{fig:desitterNF}). One of these rotating trajectories is a
fairly good approximation of the actual observed Laplace state. The
reconstruction of the dynamics is quite accurate when also a
first-order secular description of the oblate potential of Jupiter
is included in the model.

Hamiltonian normal forms obtained in this framework are prototypes
for the description of systems trapped in resonance. This de
Sitter normal form can be used to explore the dynamics both around
the librating and rotating regimes when perturbed with
higher-degree terms in the expansions of the satellite mutual
interactions, secular effects from Callisto, the Sun,
higher-degree multipoles of Jupiter, etc. A further
improvement of the model, which can be the subject of a future work,
might include dissipative perturbations due to tidal
interactions.

We illustrate the use of the 1-DOF Hamiltonian to check the
accuracy in the location of the equilibrium and how sensitive is
this when changing the nature of the perturbation.  As an
application, an inspection of the phase-plane of the normal form
allows us very easily to identify a \sl Laplace-like \rm
configuration which displays almost every feature of the observed
one.

The plan of the paper is as follows: in Section 2 we introduce the
Hamiltonian model; in Section 3 we locate the stable de Sitter
equilibrium and construct the normal form; in Section 4 we
investigate its predictive power and the sensitivity to
higher-order effects and in Section 5 we discuss possible
extensions and conclusions.

\section{An analytical model of the Laplace resonance}

We illustrate the analytic model based on the Hamiltonian method
to reconstruct the dynamics around the Laplace resonance. We
follow the standard approach of Henrard and Malhotra
\cite{He,Ma}. The model includes the most relevant interaction
and can be conveniently generalized when inserting less important
effects. Working in the Jacobi coordinate frame and considering
only planar orbits, we take into account the Newtonian monopole,
the oblateness of Jupiter and the mutual interaction of the three
satellites (Io, Europa and Ganymede) involved in the resonance.
Their osculating elements, respectively the semi-major axes,
eccentricities, mean longitudes and longitudes of peri-Jove, are
denoted as $\{a_k,e_k,\lambda_k,\varpi_k \}$ with
$k=1,2,3$.

The Newtonian monopole Hamiltonian is given by
\beq{Hkep}
H_{Kep}=-{{Gm_0m_1}\over {2a_1}}-{{G(m_0+m_1)m_2}\over {2a_2}}-{{G(m_0+m_1+m_2)m_3}\over {2a_3}}\ ,
\eeq
where $m_{0}, m_k \, (k=1,2,3) $ are, in order, the masses of Jupiter, Io, Europa and Ganymede.

The secular $J_2$-term in the oblate potential of Jupiter is
\beq{Hjup}
H_{J_{2}}=
-{{G m_0 J_2 R_{\jupiter}^2}\over {2}}\ \left[
{m_1\over a_1^3} \left(1+{3\over 2}e_1^2\right) +
{m_1+m_2\over a_2^3} \left(1+{3\over 2}e_2^2\right) +
{m_1+m_2+m_3\over a_3^3} \left(1+{3\over 2}e_3^2\right) \right] \ ,
\eeq
where $ J_2 = 1.4736 \times 10^{-2}$ is the Jupiter quadrupole coefficient in the multipole expansion of the gravitational potential and $R_{\jupiter}=71398$ kilometers is its equatorial radius. From the amount of these quantities we can state that the oblateness of Jupiter produces a very large effect, absolutely not negligible even in a first order picture: in fact the precession of peri-Joves turns out to be comparable in magnitude to the effects due to satellite couplings. 

The mutual interaction of the three resonant satellites, limiting the expansion up to first order in the eccentricities, is given by:
\beqa{Hsat1}
H_{Sat}&=&-{{Gm_1m_2}\over a_2}\left\{{B_0}\left(\alpha_{12}\right)+
\gamma_1\left(\alpha_{12}\right)e_1\cos(2\lambda_2-\lambda_1-\varpi_1)+
\gamma_2\left(\alpha_{12}\right)e_2\cos(2\lambda_2-\lambda_1-\varpi_2)\right\}\ \nonumber\\
&&-{{Gm_2m_3}\over a_3}\left\{{B_0}\left(\alpha_{23}\right)+
\gamma_1\left(\alpha_{23}\right)e_2\cos(2\lambda_3-\lambda_2-\varpi_2)+
\gamma_2\left(\alpha_{23}\right)e_3\cos(2\lambda_3-\lambda_2-\varpi_3)\right\}\ \nonumber\\
&&-{{Gm_1m_3}\over a_3}\left\{{B_0}\left(\alpha_{13}\right)\right\} + {\rm O} (e_k^2)\
, \eeqa
where the ${B_0}$ and $\gamma_k$ are defined as
\beq{gammak}
B_0(\alpha)=\frac12 {b_{1/2}^{(0)}(\alpha)} - 1, \quad
\gamma_1(\alpha)=  {1\over 2}\left(4+\alpha{d\over{d\alpha}}\right)b_{1/2}^{(2)}(\alpha),\quad
\gamma_2(\alpha)=-{1\over 2}\left(3+\alpha{d\over{d\alpha}}\right)b_{1/2}^{(1)}(\alpha)+2\alpha
\eeq
and the $b_{s}^{(j)}(\alpha)$ are the Laplace
coefficients \cite{MD}
with $\alpha_{ik},
(i,k=1,2,3)$ being $\alpha_{12}=a_1 / a_2$, $\alpha_{23}=a_2 /
a_3$ or $\alpha_{13}=a_1 / a_3$; indirect terms are included in the computation of the coefficients.

Since the eccentricities are small, it is convenient to use the modified Delaunay variables
\beq{LP}
L_i=\mu_i\sqrt{GM_ia_i}\ ,\qquad \widetilde P_i=L_i\left(1-\sqrt{1-e_i^2}\right)
\eeq
with conjugate angles $\lambda_i$ and $p_i=-\varpi_i$. In these definitions, the following combinations of masses are introduced complying with the choice of the Jacobi frame:
\beqa{}
M_1&=&m_0+m_1 \ , \\
M_2&=&m_0+m_1 + m_2\ , \\
M_3&=&m_0+m_1 + m_2 + m_3\ , \\
\mu_1&=&\frac{m_0 m_1}{M_1} \ , \\
\mu_2&=&\frac{M_1 m_2}{M_2} \ , \\
\mu_3&=&\frac{M_2 m_3}{M_3} \ . \\
\nonumber
\eeqa
Overall, we have the Hamiltonian function
\beq{HLP}
H_{A}(L_i,\widetilde P_i,\lambda_i,p_i)=H_{Kep} + H_{J_{2}} + H_{Sat}, \;  \, i=1,2,3\ .
\eeq

In order to exploit the resonances, we use the following \sl
Henrard-Malhotra \rm coordinate transformation \cite{He,Ma},
which takes into account the resonant combinations of the angles
corresponding to the Laplace resonance: \beqa{}
q_1&=&2\lambda_2-\lambda_1+p_1 \ , \quad\quad P_1 = \widetilde P_1 \ ,\label{q1}\\
q_2&=&2\lambda_2-\lambda_1+p_2 \ , \quad\quad P_2 = \widetilde P_2 \ , \label{q2}\\
q_3&=&2\lambda_3-\lambda_2+p_3 \ , \quad\quad P_3 = \widetilde P_3 \ , \label{q3}\\
q_4&=&3\lambda_2-2\lambda_3-\lambda_1 \ , \quad \;
P_4 = {\scriptstyle{\frac13}}\left(L_2-2(\widetilde P_1+\widetilde P_2)+\widetilde P_3\right) \ , \label{q4}\\
q_5&=&\lambda_1-\lambda_3 \ , \quad\quad\quad\quad \;\;
P_5 = {\scriptstyle{\frac13}}\left(3L_1+L_2+\widetilde P_1+\widetilde P_2+\widetilde P_3\right) \ , \label{q5}\\
q_6&=&\lambda_3\ , \quad\quad\quad\quad\quad\quad \;\;\;
P_6 = L_1+L_2+L_3-\widetilde P_1-\widetilde P_2-\widetilde P_3  \label{q6}
\eeqa
and, for sake of convenience, we also give the list of the old $L$-actions in terms of the new ones
\beqa{}
L_1&=&P_5-P_4-P_1-P_2 \ ,\label{L1}\\
L_2&=&3P_4+2 P_1+2P_2-P_3 \ , \label{L2}\\
L_3&=&P_6-P_5-2P_4+2P_3 \ . \label{L3}
\eeqa
In this way we obtain the transformed Hamiltonian
\beq{HLP}
H_{B}(P_{\alpha},q_a)=H_{A}\left(L_i(P_{\alpha}),P_i, \lambda_i(q_{\alpha}),p_i(q_{\alpha})\right) , \; \alpha=1,...,6, \quad a =1,...,4\  ,
\eeq
whose explicit form is given in Appendix \ref{app:A}, formula \eqref{HBE}.
Now, the relevant property of \eqref{HBE} is that $q_5$ and $q_6$ are cyclic in $H_B$, so that $P_5$ and $P_6$ are integrals of motion. Therefore we get a 4-DOF Hamiltonian system with $P_5$ and $P_6$ as parameters.

Henrard \cite{He} and Malhotra \cite{Ma}, in their respective
Hamiltonian models, proceed to expand the function around \sl
reference \rm values for $P_4,P_5,P_6$, noticing that they are
intrinsically large quantities when compared with $P_1,P_2,P_3$.
To preserve the accuracy in the following computations, we prefer
at the moment not to perform this expansion. However, we will
exploit these reference values as \sl seeds \rm for the
computation of the equilibria. Moreover, since it is large but not
conserved, we split the momentum $P_4$ according to
\beq{M4}
P_4=\overline P_4 + \delta P_4 \ ,
\eeq
where $\overline P_4$, defined below in \eqref{IM1},
is a fixed value and $\delta P_4$ denotes a small variation. The 4-DOF system generated by
$H_B$ is still quite complicated (and most probably
non-integrable). We proceed by investigating its equilibria and
the dynamics around them.

\section{The de Sitter equilibrium}

We recall that the Laplace resonance is a three-body orbital resonance involving the Jovian moons described by the relations
\beqa{res}
2\lambda_2-\lambda_1-\varpi_1&=&0\ ,\nonumber\\
2\lambda_2-\lambda_1-\varpi_2&=&\pi\ ,\nonumber\\
2\lambda_3-\lambda_2-\varpi_2&=&0\ .
\eeqa
As a consequence, the resonant Laplace argument $\phi_L$, defined by
$$
\phi_L\equiv \lambda_1-3\lambda_2+2\lambda_3\ ,
$$
librates around $\pi$: this implies that there can never be a triple conjunction
of Io, Europa and Ganymede.

On the other side, the de Sitter equilibrium \cite{Si,BH}
corresponds to the set of relations: \beqa{}
q_1&=&2\lambda_2-\lambda_1-\varpi_1 =0 \ ,\label{q1ds}\\
q_2&=&2\lambda_2-\lambda_1-\varpi_2 =\pi \ , \label{q2ds}\\
q_3&=&2\lambda_3-\lambda_2-\varpi_3 =\pi \ , \label{q3ds}\\
q_4&=&3\lambda_2-2\lambda_3-\lambda_1 (= -\phi_L)=\pi  \ . \label{q4ds}\\
\nonumber
\eeqa
We remark that these relations identify only one particular equilibrium state and that, as a matter of fact,
several other equilibrium solutions are possible, each of them belonging to the de Sitter family (see \cite{BH}).
By denoting with the sequence $\{0,\pi,\pi,\pi\}$ the values corresponding to \eqref{q1ds}--\eqref{q4ds},
we can see that the combinations involving respectively one, two or three exchanges
{between} $0$ {and} $\pi$, are still equilibria. 
Moreover, a rotation of the whole system by $\pi$,
corresponding to an overall exchange of each $0$ with $\pi$ and of
each $\pi$ with $0$, provides an additional set of eight
equilibria. We have therefore a collection of 16 possible de
Sitter equilibria. However, as found by Hadjidemetriou and Michalodimitrakis \cite{HM81} and analytically checked in the works by Broer and
Han{\ss}mann \cite{BH} and Broer and Zhao \cite{ZHB}, all of
them are linearly unstable, but for the one identified by
\eqref{q1ds}--\eqref{q4ds} and its rotated counterpart. To this
linearly stable configuration and the corresponding equilibrium value of the momenta, we will henceforth refer as \sl the
de Sitter equilibrium. \rm
We remark that these last equilibrium values are slightly displaced when adding
to the Hamiltonian \eqref{HLP} a small perturbation sharing its symmetries
with respect to the angles. Since the precise evaluation of these displacements can be
difficult, we will generically have small librations around the given reference
values of the combination angles.

Concerning the de Sitter equilibrium, the essential difference
with respect to the standard Laplace resonant configuration is
that, in addition to the conditions in \eqref{res}, we now also
have \eqref{q3ds}, that is the libration of $q_3$ around $\pi$.
Since observations indicate that $q_3$ is rotating \cite{ShMa},
we realize that de Sitter equilibrium is not the
observed configuration of the Galilean satellites. However, we can study
under which conditions the real system fails to stay in this
status.

To locate the equilibrium, we look for the solution of the system
of equations \beq{DHDS} {{\partial
H_{B}(P_{\alpha},q_{a})}\over{\partial P_{b}}} \bigg\vert_{DS} =0,
\quad\quad {{\partial H_{B}(P_{\alpha},q_{a})}\over{\partial
q_{b}}} \bigg\vert_{DS} =0\  , \; \alpha=1,...,6, \quad a,b
=1,...,4 \ , \eeq where the suffix \sl DS \rm means that the
derivatives are computed at the values in
\eqref{q1ds}--\eqref{q4ds}. An exact general solution is difficult
to get explicitly; most probably it could be obtained if $H_B$ is
expanded as in \cite{He} and \cite{Ma} and retaining linear terms
in the small momenta. We instead proceed by inserting the explicit
values of the numerical parameters and exploiting the \tt FindRoot
\rm function of {\sc mathematica}\textregistered .

The values of the parameters are determined according to the following conventions.
We scale physical units by
choosing $Gm_0=1$ and $a_1=1$. Actually, this last choice is
equivalent to scale distances by the semi-major axis of a \sl
virtual \rm Io, which is taken as the nominal mean value coming
from ephemerides. Analogously, the average observed values of
$a_2,a_3$ and of the three eccentricities, are used to initialize
the computations:  in Table \ref{T1} we list the values of the
elements used in \cite{La-A} taken from the IMCCE, which will be
used in the forthcoming computations. The true proper\footnote{With the term \sl proper \rm we refer
to values which are produced by the analytical computation.}
semi-major axes involved in the dynamical status are among the outcome of the
procedure and will slightly differ from them. The values of the
Laplace coefficients used in the model are those corresponding to
the mean elements: in Table \ref{T2} we list their values for the
ephemeridal figures of Table \ref{T1} (denoted as [LDV04]) and, for sake of comparison,
with the ephemerides extracted from the NASA's SPICE
toolkit\footnote{Ephemerides from the IMCCE (available from:
\tt http://www.imcce.fr/ephemeride.html\rm) correspond to Dec 25th
1982; those from NASA correspond instead to Jan 1st 2000 (Spice can be downloaded at
\tt https://naif.jpl.nasa.gov/naif/toolkit.html\rm)} (denoted as [J2000]) and also with the values
corresponding to the \sl exact \rm commensurability (last row),
$\alpha_{12}=\alpha_{23}=2^{-2/3}$. The values of
${B_0}(\alpha_{13})$, not included in Table \ref{T2}, are
all very close (within the last displayed digit) to the common
value $0.04266$.

\begin{table}
    \begin{tabular}{@{}lrrr@{}}
  \hline
  \hline

$\rm{Element}$                                              &\rm{Io}                        & \rm{Europa}                   & \rm{Ganymede} \\
\hline
 semi-major axis [{\rm km}]                                 &  422030.686                   & 671262.329                &       1070622.862    \\
 eccentricity                                                       &  0.004165                     & 0.009366                      &       0.001500  \\
 orbital period \, [{\rm d}]                                    &  1.769137                     & 3.551182                  &   7.154554  \\
 mean motion \, [$^{\circ}$/d]                              &  203.4890                     & 101.3747                      &   50.3176  \\
 mass ratio \, [$m_k/m_0$]                                  &  $4.70\times10^{-5}$      & $2.56\times10^{-5}$       &   $7.84\times10^{-5}$  \\
\hline
\end{tabular}
\caption{Mean orbital elements according to Lainey et al.
\cite{La-A}} \label{T1}
\end{table}

Using these values in the definitions \eqref{LP} and in the
transformed momenta (\ref{q4})--(\ref{q6}), we compute the
following \sl seeds \rm for $\overline P_4$ and the integrals of
motion $P_5, P_6$: \beq{IM1} \overline P_4 = 0.2289560, \quad
P_5=1.2289716, \quad P_6=4.3437055 \ .
\eeq
$\overline P_4$ is the
value of $P_4$ for the ephemeridal elements. In this way, solving
the system \eqref{DHDS} we get the following solution for the de Sitter equilibrium:
\beqa{}
P_1^*&=&9.74805159\times10^{-6} \ , \label{P_1ds}\\
P_2^*&=&3.55566494\times10^{-5} \ , \label{P_2ds} \\
P_3^*&=&6.84513718\times10^{-7} \ , \label{P_3ds}\\
\delta P_4^*&=&-7.29208631 \times10^{-6} \ .  \label{P_4ds}
\eeqa
From these values and those in \eqref{IM1}, using (\ref{L1})--(\ref{L3}) we also get
\beqa{}
L_1^*&=&0.99997729 \ , \label{L1ds}\\
L_2^*&=&0.68693684 \ , \label{L2ds} \\
L_3^*&=&2.65683739 \ , \label{L3ds}
\eeqa
from which, using again their definitions in \eqref{q4}--\eqref{q6}, we can recompute the values of $P_4,P_5,P_6$ for consistency. The corresponding proper orbital elements are as follows:
\beqa{}
a_1^*&=&1.0000016 \ , \label{a1ds}\\
a_2^*&=&1.5905229 \ , \label{a2ds} \\
a_3^*&=&2.5368556 \ , \label{a3ds} \\
e_1^*&=&0.0044155 \ , \label{e1ds}\\
e_2^*&=&0.0101746 \ , \label{e2ds} \\
e_3^*&=&0.0007178 \ . \label{e3ds}
\eeqa
Comparing these values with the figures of Tables \ref{T1} and \ref{T2}, we see that the new semi-major axis ratios are still very close to the original ones and that the only slight but substantial change is in the eccentricity of Ganymede.

\begin{table}
    \begin{tabular}{@{}lccc@{}}
  \hline
  \hline

\rm{semi-major axis ratio}                                      &${B_0}$        & $\gamma_1$    & $\gamma_2$ \\
\hline
 $\alpha_{12}=0.628712$ \cite{La-A}                         &  0.129531             & $-1.183999$       &       0.424936    \\
 $\alpha_{12}=0.628733$ [J2000]                             &  0.129540           & $-1.184091$       &       0.424985    \\
 $\alpha_{23}=0.626983$ \cite{La-A}                         &  0.128561            & $-1.174833$       &       0.420078   \\
 $\alpha_{23}=0.627044$ [J2000]                             &  0.128600             & $-1.175330$       &       0.420341    \\
 $\alpha=0.629961=2^{-2/3}$                                     &  0.130217             & $-1.190494$       &       0.428390    \\
 \hline
\end{tabular}
\caption{Laplace coefficients.}
\label{T2}
\end{table}

\begin{figure}[h]
\centering
\hglue-1cm
\includegraphics[width=10truecm]{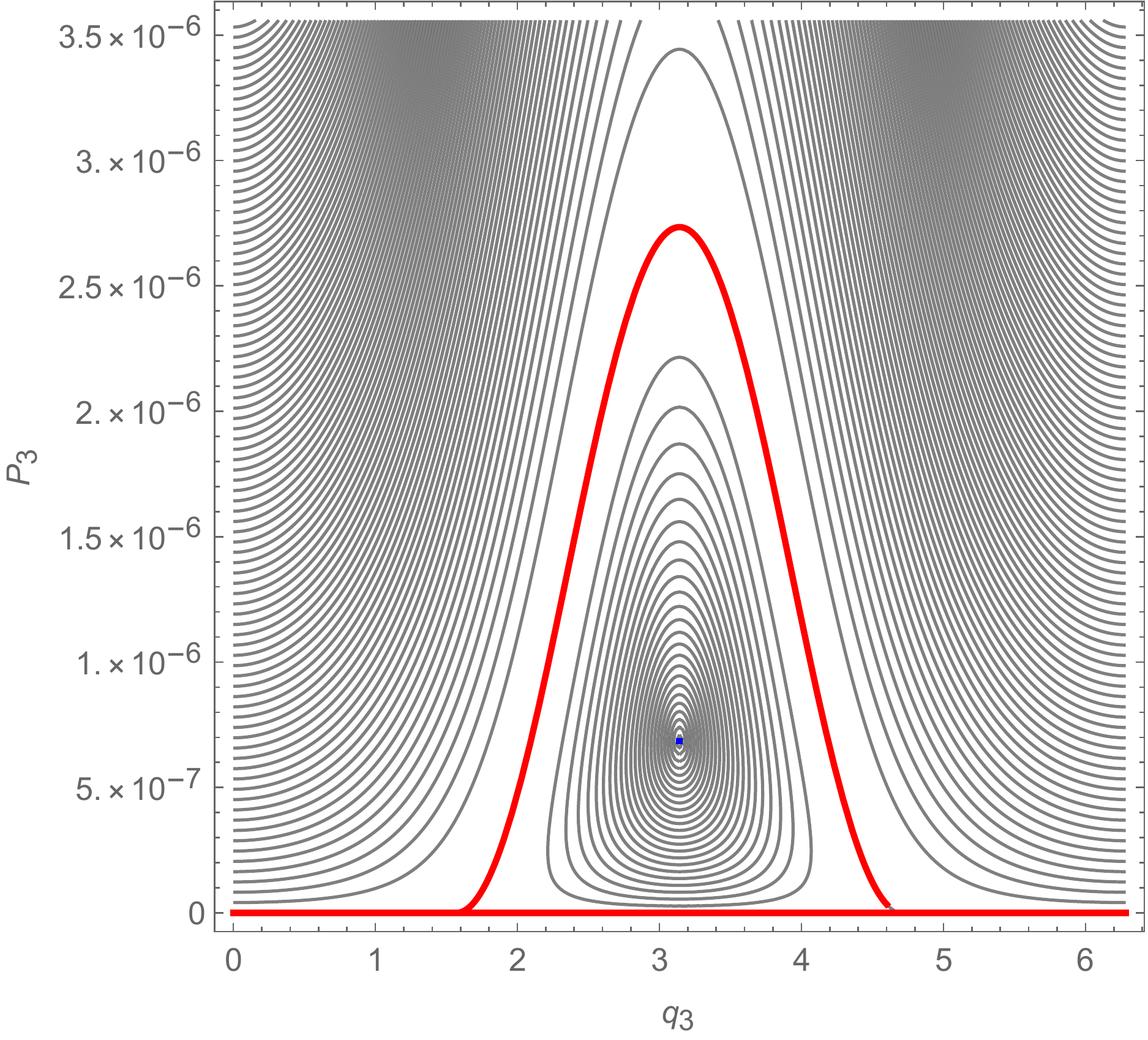}
\caption{The normal form around the de Sitter equilibrium: the red curve represents the boundary which delimits the libration domain.}\label{fig:desitterNF}
\end{figure}

A normal form Hamiltonian can be constructed to describe the
reduced dynamics around the de Sitter equilibrium, in much the
same way as semi-analytical constructions have been performed to
get expansion solutions \cite{Br}, making use of the advantages
of the Lie-transform normalization \cite{Kamel}. In order to
apply the Lie-transform algorithm \cite{gior,cefthym}, we now
expand the function $H_B$ around the solution values $P_a^*$ and
keep terms up to second order in the momenta. This choice may at first
seem to conflict with the first order expansion in the
eccentricities. However, it is required to properly account for
the relative importance of low-order non-resonant terms. The procedure to compute the normal form closely resembles the approach
followed by Henrard \cite{He}: a sequence of near-identity canonical transformations is performed
in order to remove selected terms in agreement with the specific aims
of the normal form. Henrard's aim was to leave only the Laplace argument and therefore
to eliminate $q_k, k=1,2,3$; instead, we are interested in the dynamics of the $(P_3,q_3)$ degree of freedom
and therefore we eliminate the angles $q_1,q_2$ and $q_4$.
Therefore, by using the equilibrium values \eqref{P_1ds},
\eqref{P_2ds} and $P_4^*=0.22894897$ computed from \eqref{q4} and
\eqref{P_4ds}, we
get a 1-DOF Hamiltonian depending on $P_3$ and $q_3$ with explicit
form
\beq{HNF}
H_{DS}(P_3,q_3)=-0.003728 P_3 - 1.6495 P_3^2 -
(6.1351\times10^{-6} \sqrt{P_3} -
 1.1546\times10^{-5} P_3^{3/2} ) \cos q_3 \ .
 \eeq
The contour plot of this function, which is strictly related
to the \sl second fundamental resonance model \rm introduced in \cite{HL},
is displayed in Fig.\ref{fig:desitterNF}. The red curve represents the boundary which delimits the libration domain around the de Sitter equilibrium. The Laplace configuration is outside this domain, where the dynamics is rotational and $q_3$ ranges from 0 to $2 \pi$. In this way we get a simple tool to highlight the reduced phase-space structure around the 4:2:1 resonance.

\section{Integrations with the Hamiltonian model}

We can exploit the plot of Fig.\ref{fig:desitterNF} to investigate both the de Sitter stable libration and the actual Laplace state by locating suitable initial conditions.

Concerning the de Sitter equilibrium, we can at first ask two
important questions: how precisely it is located by exactly evaluating the
values of $P_a^*$ and how sensitive
is this location with respect to higher-order effects. The first
question arises from the fact that our computation is only in
principle \sl exact: \rm two factors of uncertainties are given by
the finite accuracy of the root-finding process and by the small
inconsistency due to the fact that the Laplace coefficients used in the
computations do not coincide with those corresponding to the
elements of the solutions. The second question is instead more
generally related to the global persistence of the equilibrium
with respect to the reintroduction of all other perturbing effects
excluded in the simplified model.
\begin{figure}[h]
\centering
\hglue-1cm
\includegraphics[width=8truecm]{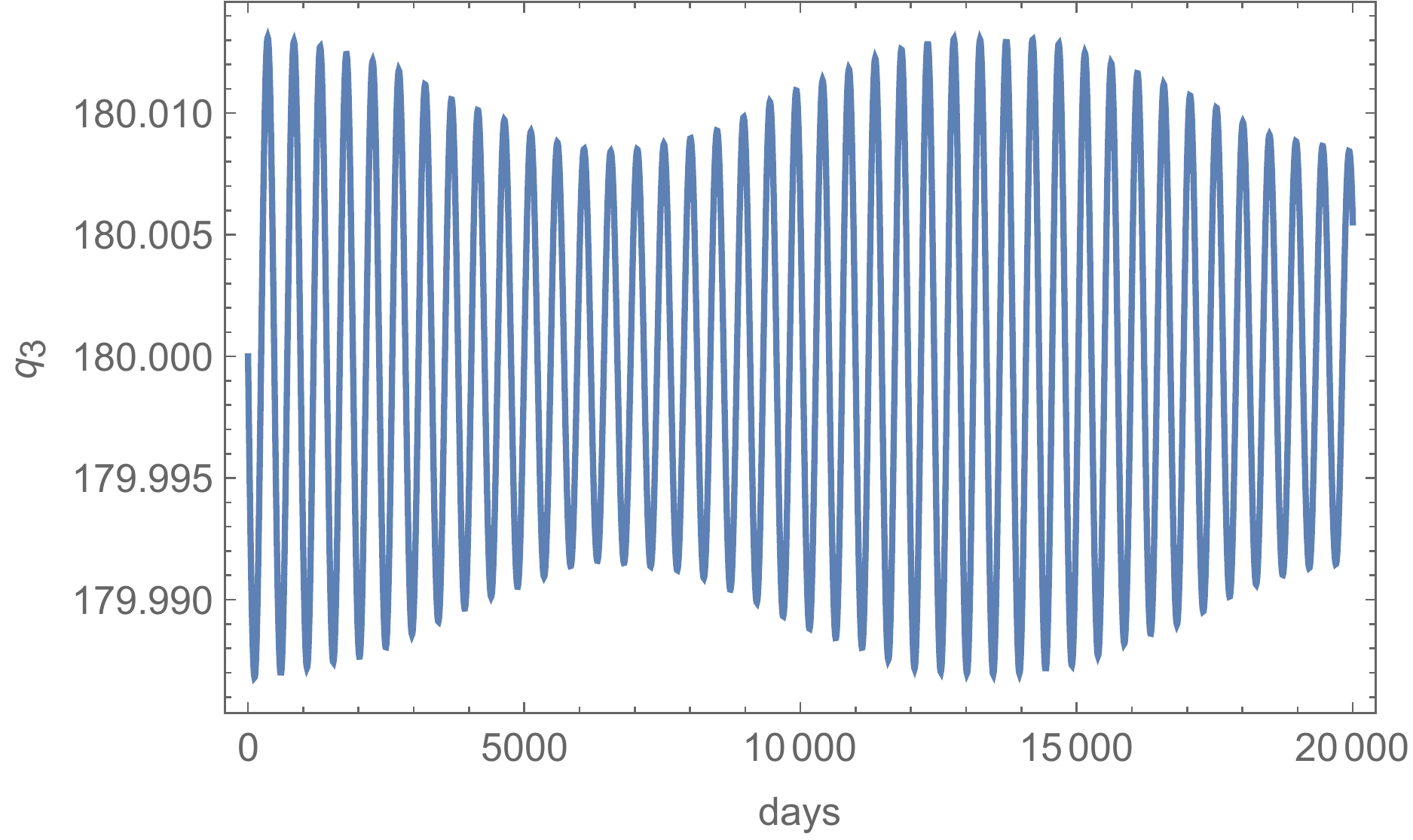}
\includegraphics[width=8truecm]{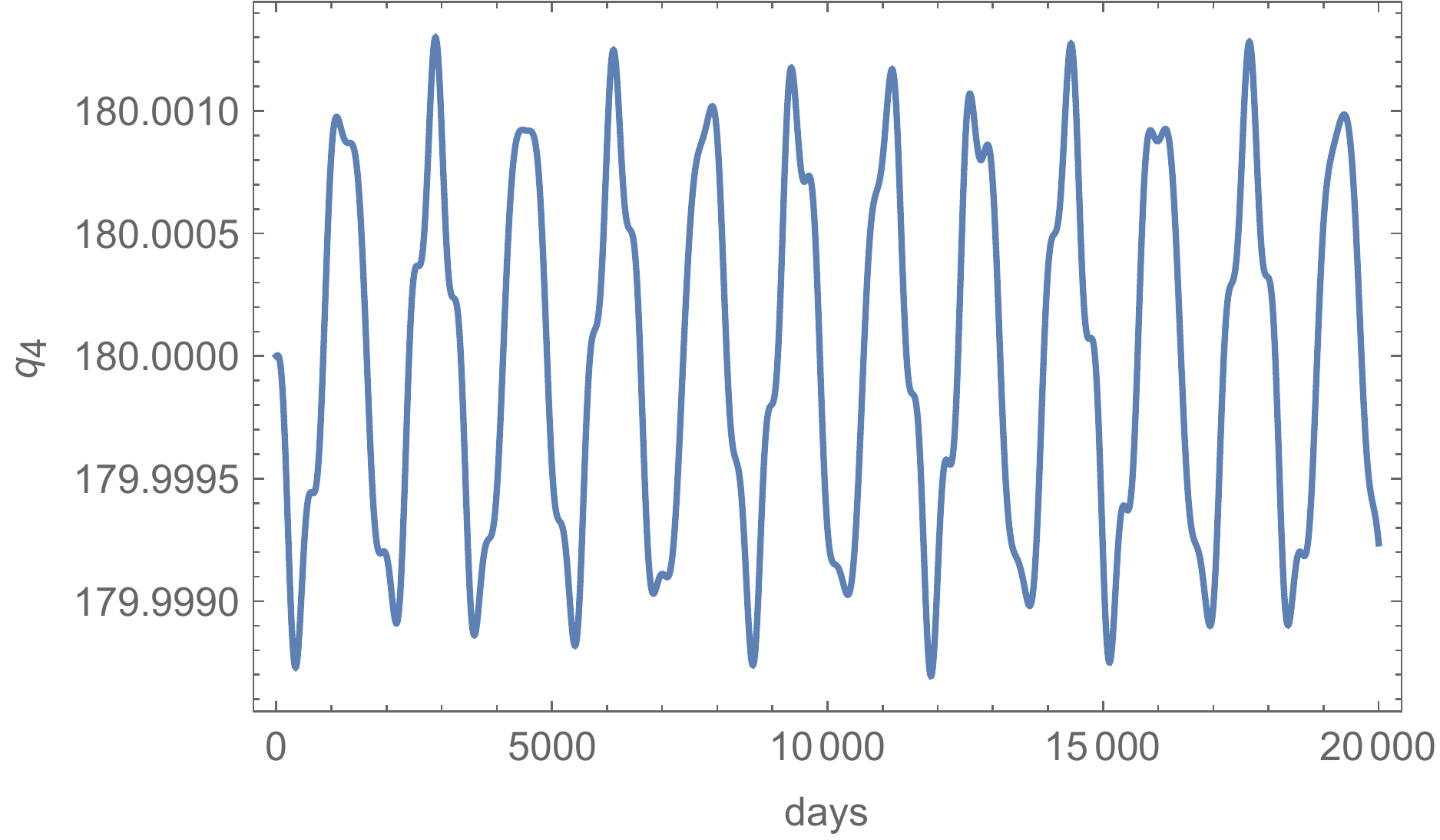}
\caption{Uncertainty in the location of the de Sitter equilibrium: left, the libration of the $q_3$
combination; right, the libration of the Laplace angle (both angles are in degrees).}
\label{fig:ds1}
\end{figure}

In Fig.\ref{fig:ds1} we see the numerical solutions of the canonical equations given by the Hamiltonian $H_B$ with initial conditions corresponding to the de Sitter equilibrium.
The fluctuations of the $q_3$ combination angle correspond to an uncertainty of order $10^{-4}$, that of the Laplace angle to $10^{-5}$. From the simulations,  also the equilibrium values of the momenta $P_a, a=1,...,4,$ fluctuate of the same amount ($10^{-4}$), whereas the integrals $P_5,P_6$ are exactly conserved. These results are in substantial agreement with the expected variations of the values of the Laplace coefficients (cfr. Table \ref{T2}): one could think to a refinement of the procedure with an iterative approach using updated values of the coefficients at each step.

Concerning higher-order effects, we have performed some investigations by extending the model with the addition of the following terms:

$\bullet$ the satellite self-interactions up to 2nd order in the eccentricities;

$\bullet$ the secular influence of Callisto and the Sun;

$\bullet$ the octupolar term in the expansion of the field of Jupiter.
\begin{figure}[h]
\centering
\hglue-1cm
\includegraphics[width=8truecm]{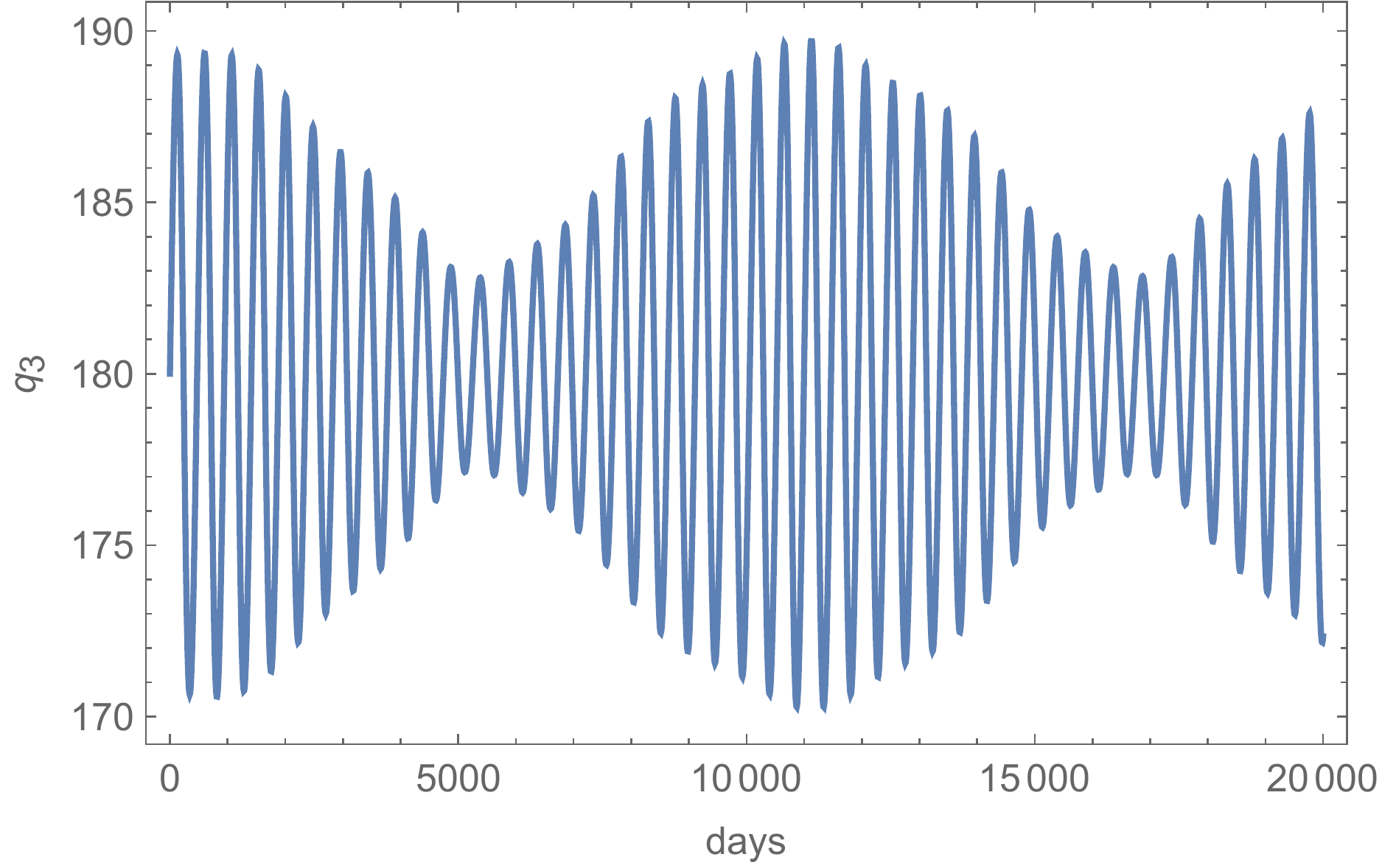}
\includegraphics[width=8truecm]{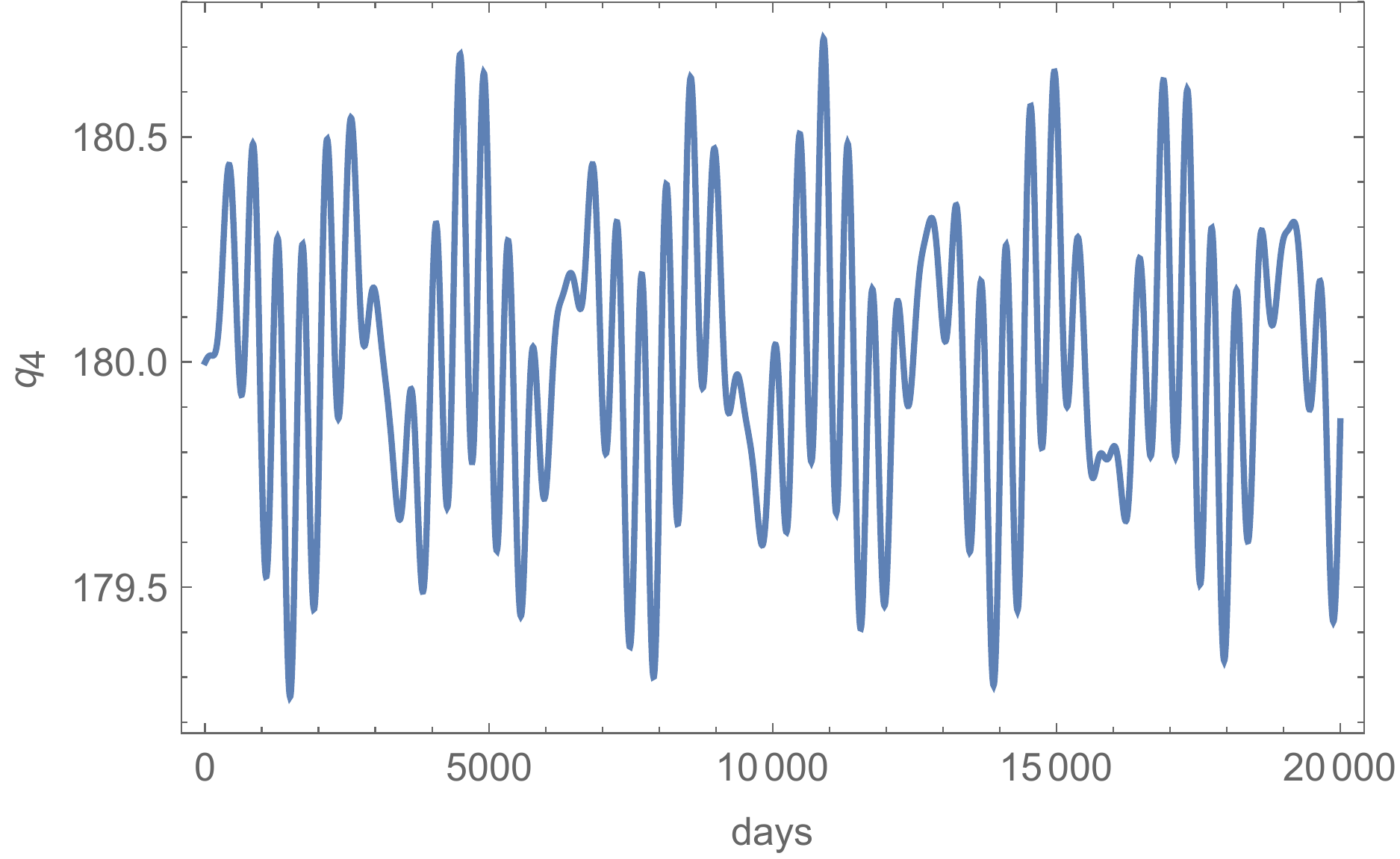}
\caption{Persistence of the de Sitter equilibrium with respect to additional high-order
effects (see text): left, the libration of the $q_3$ combination; right, the libration of the Laplace angle.}
\label{fig:ds2}
\end{figure}

The expressions of these higher-order terms are written in Appendix \ref{app:B}. In Fig.\ref{fig:ds2} we see the numerical solutions of the canonical equations given by the second-order model Hamiltonian including the above effects starting with the same initial conditions as before. We see that the $q_3$ angle shows appreciable quasi-periodic oscillations, but well within the limit of the libration area; the Laplace angle has still a quite small amplitude of variation. Three time-scales emerge from the plots in Figg.\ref{fig:ds1}-\ref{fig:ds2}: a 485-days oscillation related to the 2:1 two commensurabilities (see the frequency value of \eqref{dia2} below), a $\sim 2000$-days oscillation in the Laplace argument and a low-frequency modulation, probably a beat between the two resonances. 

Also in the \sl real \rm Laplace resonance, a very small amplitude
of the librations of $q_1,q_2$ and $q_4$ is usually reported \cite{YoPe}. We
can conjecture that this status is in some sense close to that of
the de Sitter equilibrium but not trapped at it. A reasonable
assumption is that of choosing initial conditions for motion of
the model system of the Hamiltonian $H_{B}(P_{\alpha},q_{\alpha})$
given by $P_1=P_1^*,P_2=P_2^*,P_4=P_4^*$ for the actions and
$q_1=0,q_2=\pi,q_4=\pi$ for the angles not directly involved in
the rotation. The remaining two variables, $P_3$ and $q_3$ can be
initialized at any point outside the boundary curve of
Fig.\ref{fig:desitterNF}.

\begin{figure}[h]
\centering
\hglue-1cm
\includegraphics[width=10truecm]{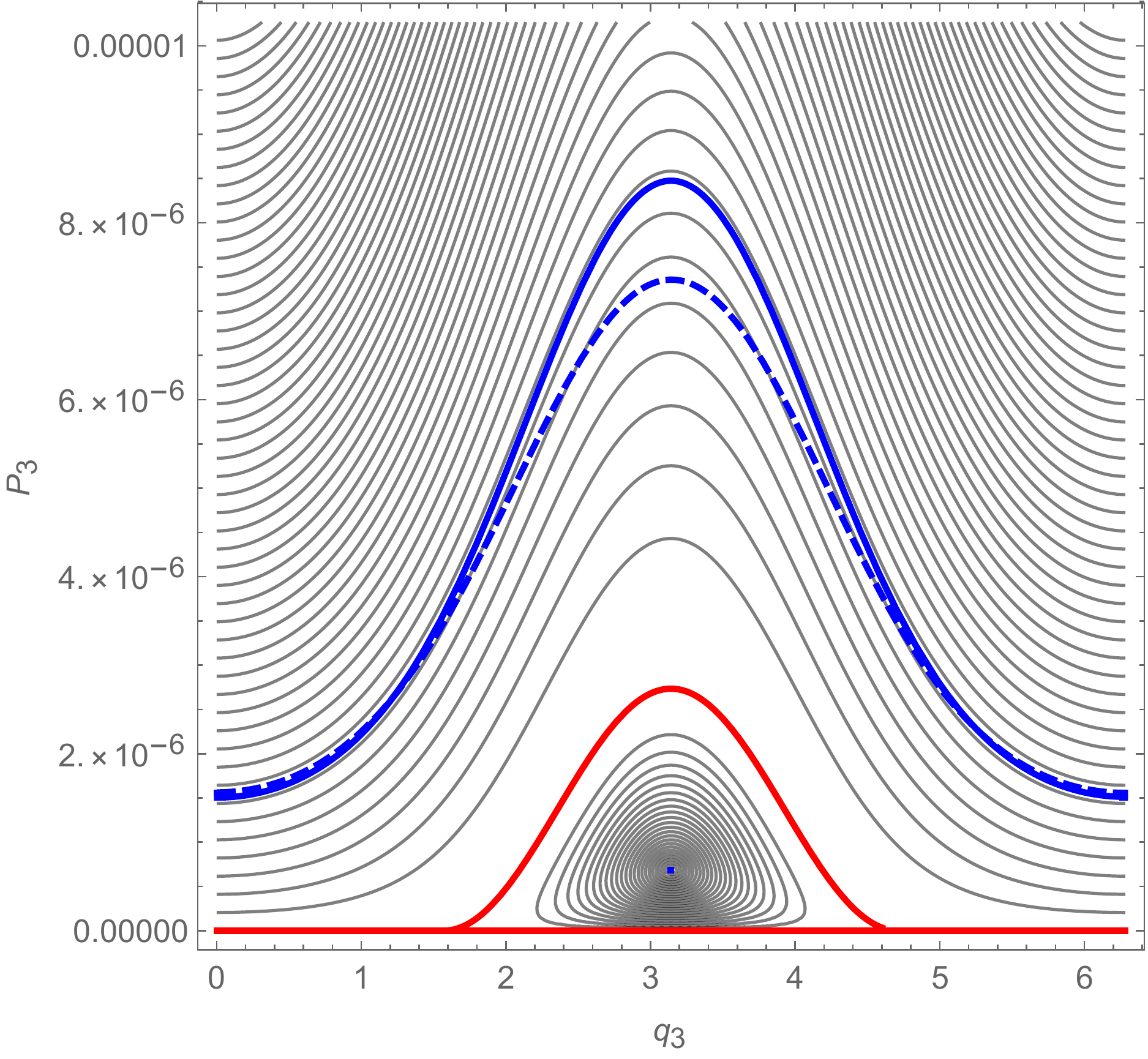}
\caption{The Laplace resonance in the reduced phase-plane: the blue curves are the projections of
the numerical integrations with the simplified model of Hamiltonian $H_B$ (continuous curve) and the
2nd-order model (dashed curve).}
\label{fig:desitterNFS}
\end{figure}

Here we report some results obtained with the choice
$$P_3=1.5 \times 10^{-6}, \quad q_3=0 \ ,$$
solving the equations of motion both in the standard model given
by  $H_B$ and in the \sl second-order model \rm (satellites at 2nd
order in the $e_k$ + \sl octupolar \rm Jupiter + Callisto + Sun).
The solutions in the $(q_3,P_3)$-plane are shown in
Fig.\ref{fig:desitterNFS}: the blue curves are the projections of
the phase evolutions. The continuous curve is the result for the
simplified model of Hamiltonian $H_B$, the dashed curve is the
outcome of the integration with the same initial conditions but in the
2nd-order model. We can see that the former curve practically
coincides with a level curve of the normal form Hamiltonian
function of \eqref{HNF} confirming that $H_{DS}$ is a very good
integrable approximation of the resonant dynamics. Moreover, the
second curve is only slightly displaced form the first, testifying
the fact that $H_B$ captures the dominant effects and that the
additional ones due to the more complete model have only quite small
consequences.

\begin{figure}[h]
\centering
\hglue-1cm
\includegraphics[width=8truecm]{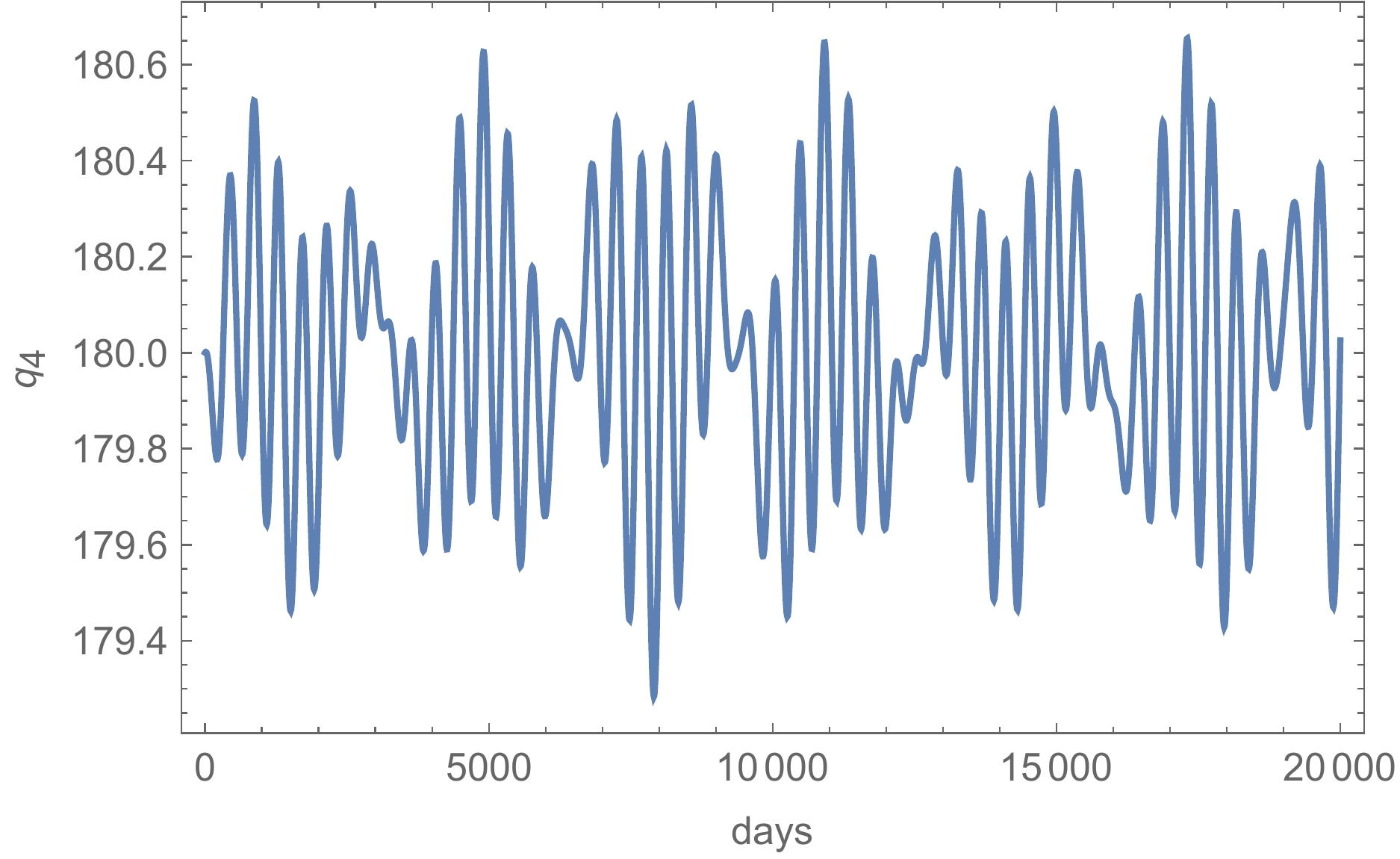}
\includegraphics[width=8truecm]{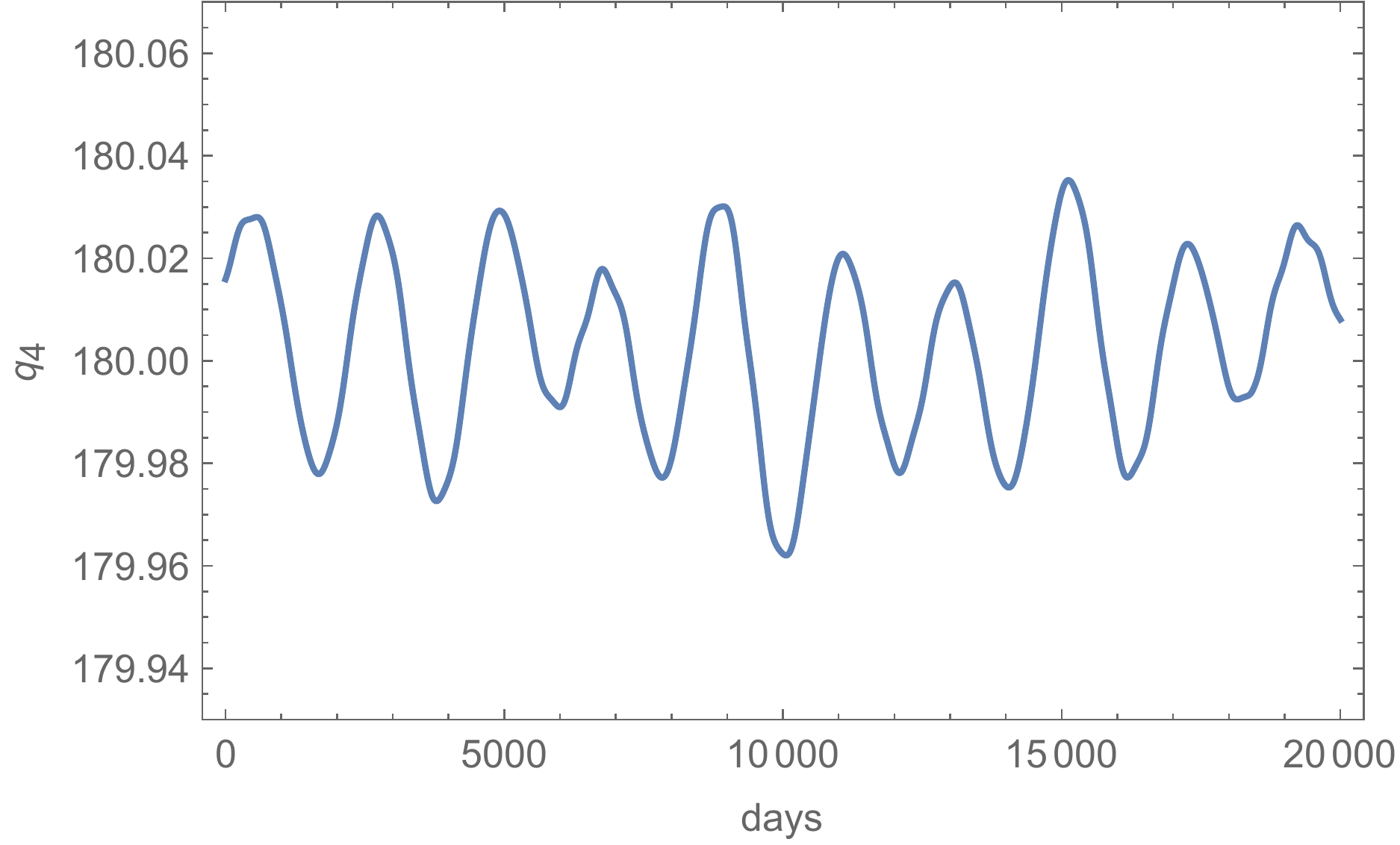}
\caption{The Laplace angle: unfiltered data on the left; filtered data on the right.
Days on the time-axis and angles in degrees.}
\label{fig:lapla}
\end{figure}

In this case the ensuing mean motions are
$$ n_1 = 203.4847^{\circ}/day, \quad n_2 = 101.3746^{\circ}/day, \quad n_3 = 50.3196^{\circ}/day \ ,$$
which are reasonably close to the observed values (see Table \ref{T1}), even if the resonant combinations
\beq{dia2} n_1 - 2 n_2 = n_2 - 2 n_3 = 0.7355^{\circ}/day \eeq
turn out to be slightly different from the observed value $
0.7395^{\circ}/day $ \cite{Br}. However, we can see from
Fig.\ref{fig:lapla} that the Laplace angle has a quite small
amplitude of libration: in the left panel we see the data as they
are produced in the integration; in the right panel, the data are
filtered with a time constant of 1000 days, so to allow a
comparison with the fully-numerical solutions of Musotto et al.
\cite{MuVaMoSc}. Also the libration period of about 2070 days seems to
be almost correctly predicted.

\begin{figure}[h]
\centering
\hglue-1cm
\includegraphics[width=8truecm]{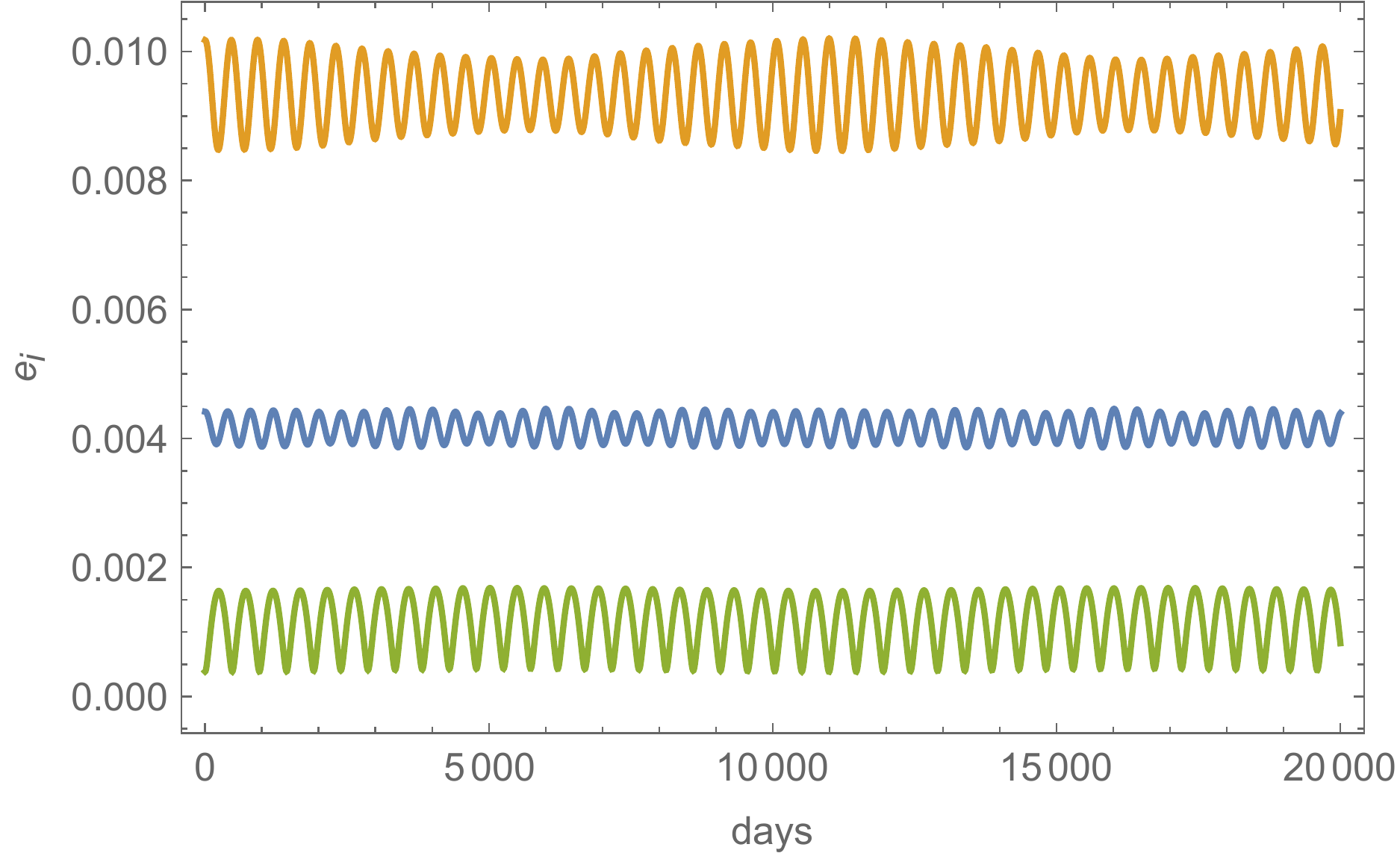}
\includegraphics[width=8truecm]{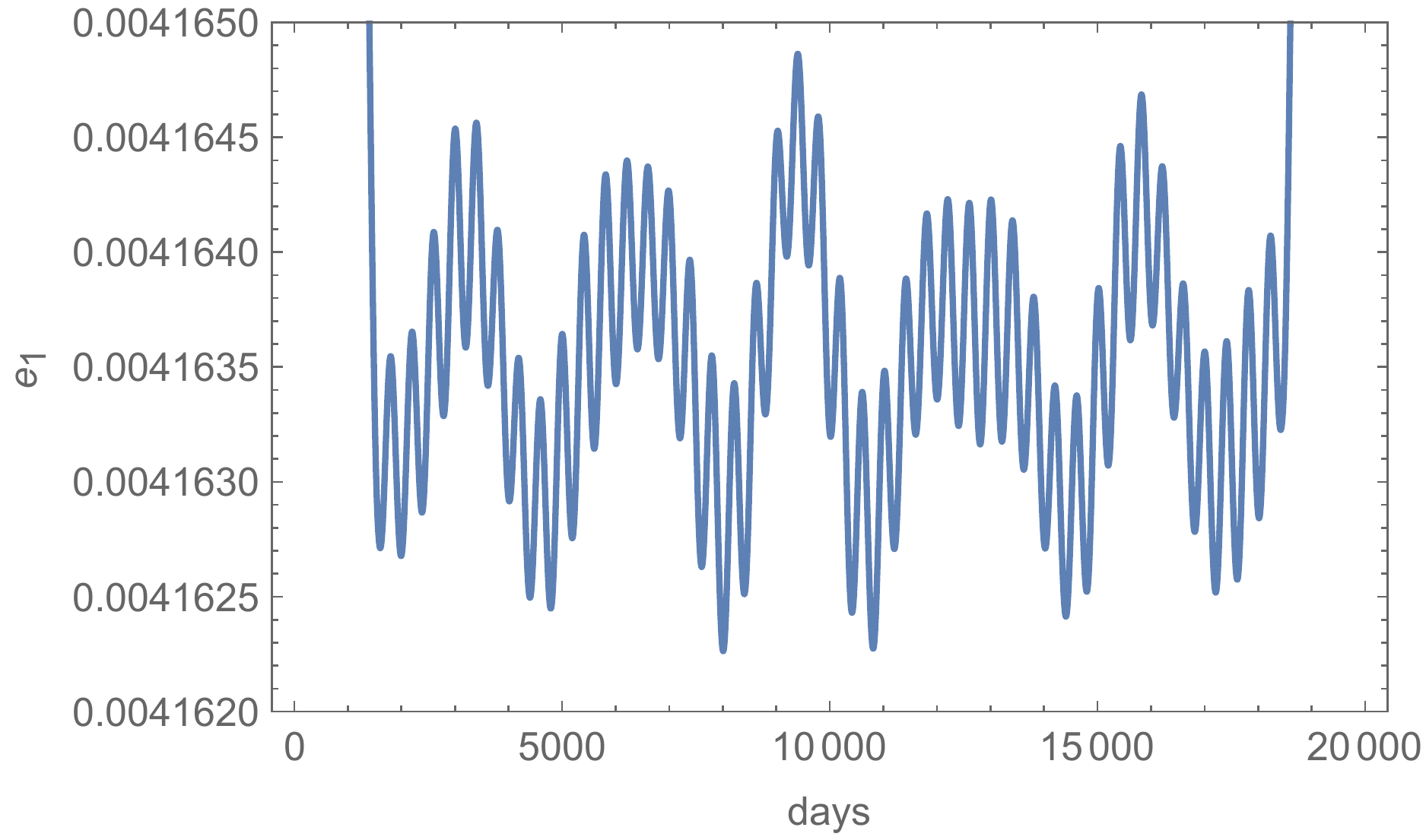}
\caption{The eccentricities: in the left panel are plotted the eccentricities of Io (blue),
Europa (orange) and Ganymede (green); in the right panel is plotted the filtered eccentricity of Io.
}\label{fig:eccen}
\end{figure}

In Fig.\ref{fig:eccen} we report the corresponding results concerning the
eccentricities. In the left panel, the three time evolutions of the eccentricities of Io, Europa and Ganymede are plotted as they are produced by the integrator. In the right panel, the data on the eccentricity of Io are filtered still with a time constant of 1000 days (the uprising curves are an artifact of the finite-time filtering operation).

\section{Conclusions}

We have presented a simple model of the resonant interaction of
the first three Galilean satellites of Jupiter. The back-bone of
the Hamiltonian dynamical system associated with the model, based
on the Keplerian and quadrupolar field of Jupiter and the first
order self-interactions of the satellites, is captured by a 1-DOF
normal form that permits to distinguish in a clear way the
librational regime around the de Sitter equilibrium and the
rotational one associated with the observed Laplace resonance.

We have investigated the uncertainties in the location of the de
Sitter equilibrium, its sensitivity to higher-order perturbations
and how close is the dynamics of the libration to the values which roughly correspond to the actually
observed state. Looking at Fig.\ref{fig:desitterNFS}, we can heuristically deduce that, starting with
a value of $P_3$ which is chosen so to reproduce realistic semi-axes and eccentricities (notwithstanding the limitations of the model), the interval of initial values $q_3 \in (135^{\circ},225^{\circ})$
corresponds to librating rather than rotating trajectories. This gives a measure of the difference between the solutions associated to the de Sitter and the Laplace states.

\begin{figure}[h]
\centering
\hglue-1cm
\includegraphics[width=8truecm]{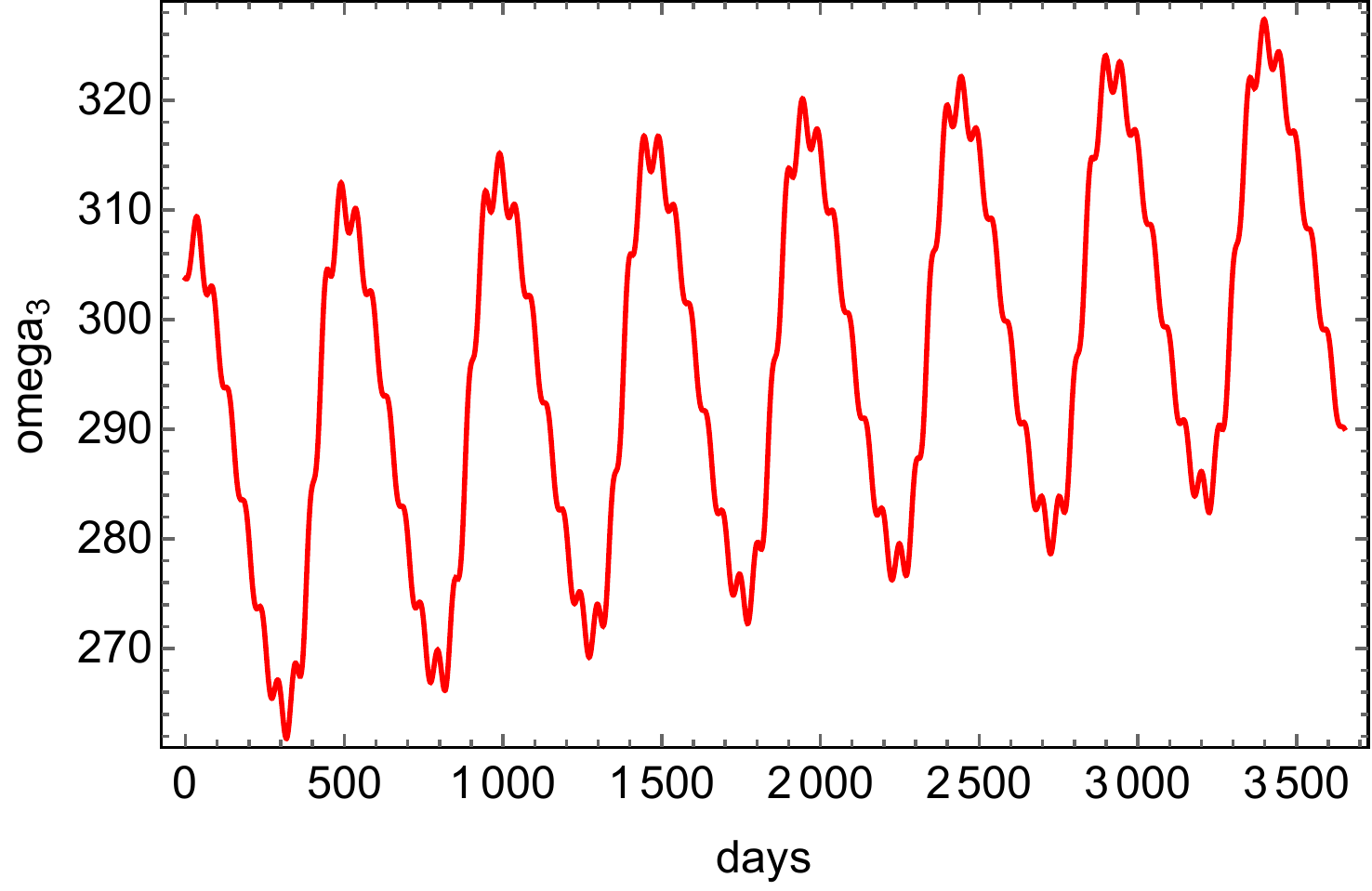}
\includegraphics[width=8truecm]{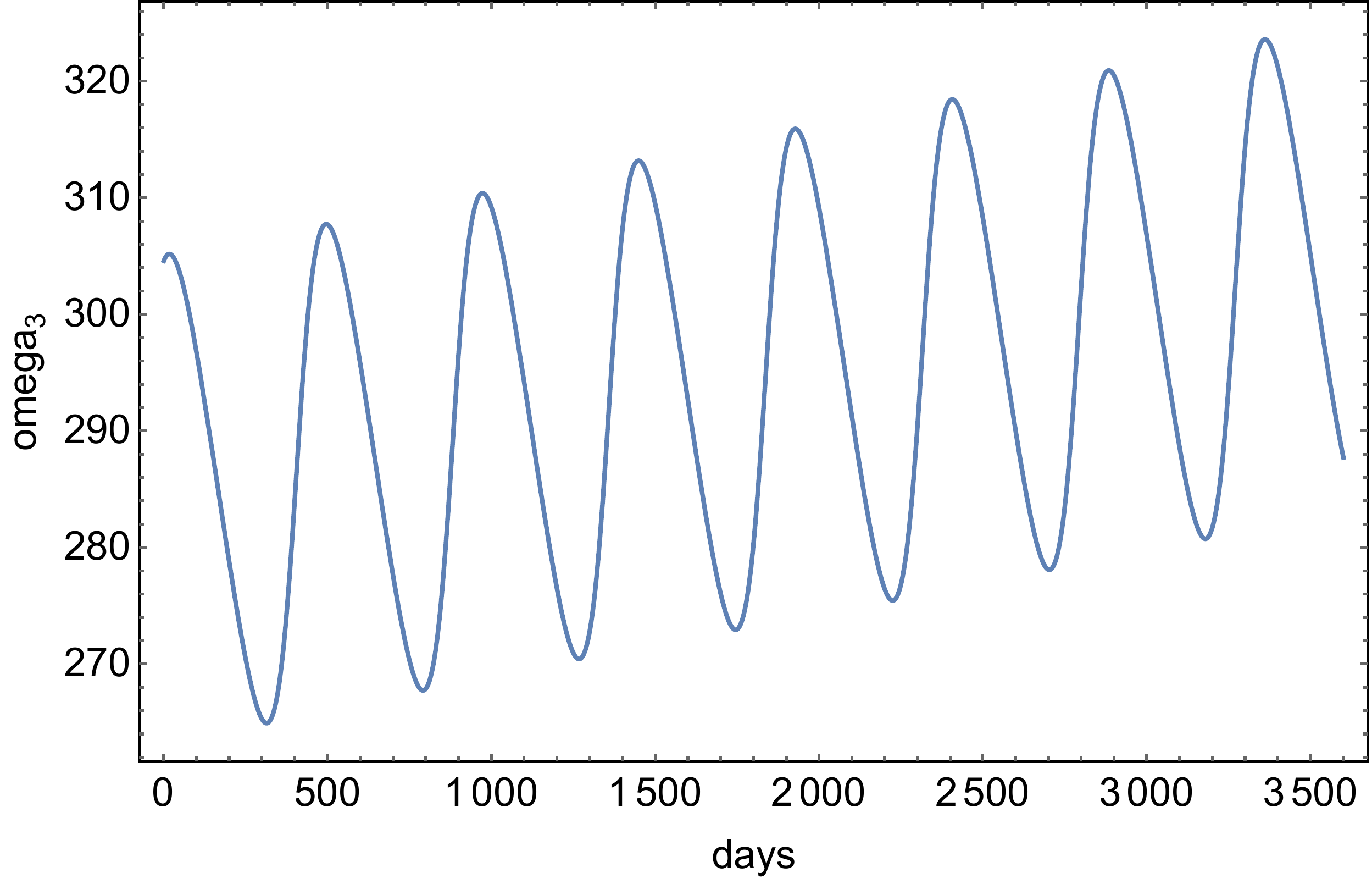}
\caption{The apsidal precession of Ganymede: in the left panel is plotted the peri-jove as given by the ephemerides; in the right panel is plotted the outcome of the integration with the first-order model.
}\label{fig:g3eph}
\end{figure}

A possible observational
test concerns the apsidal precession of Ganymede. In fact, from \eqref{q3ds}, since longitudes advance almost linearly in time, $\varpi_3$ should be  modulated at the same 485-days period of the 2:1 commensurability,
with an amplitude spanning $\sim 20^{\circ}$. We have checked this claim by using ephemerides (see the left panel of Fig.\ref{fig:g3eph}) and verified with the outcome of the integration with our first-order model (see the right panel of Fig.\ref{fig:g3eph}).

We conclude by mentioning that dissipative effects, like those due to tidal forces might induce an adiabatic variation of the parameters. The inclusion of such effects can lead to conclusions about
a possible capture or escape from the exact resonance \cite{HL,Ma}. This topic is certainly interesting and will be the object of a future study. 
It would also be interesting to apply the results of this work to test the nature of
the Laplace resonance detected in extra-solar multi-planetary systems \cite{BDH,MCB,Pa16}, possibly extending the pioneering analysis
made by Malhotra \cite{Ma} to
the cases in which the 2:1 commensurabilities are not in the standard 1:1 ratio.

\section*{Acknowledgements}

We deeply thank Sylvio Ferraz-Mello, Heinz Han{\ss}mann, Luciano
Iess and Renu Malhotra for very fruitful discussions. We also
acknowledge the Italian Space Agency for its support through the grant 2013-056-RO. Finally, we extend our gratitude to GNFM/INDAM.





\vfill\eject

\appendix

\section{Complete Hamiltonian in Henrard-Malhotra coordinates}\label{app:A}

The complete form of the Hamiltonian \eqref{HLP} including the Keplerian, the quadrupole field and the mutual interactions at first order in the eccentricities is:
\beqa{HBE}
H_{B}(P_{\alpha},q_a)
&=&-\frac{m_1}{2(-P_1 - P_2 - P_4 + P_5)^2 (1 + \epsilon_I)} \nonumber\\
&  &-\frac{m_2^3 (1 + \epsilon_I)^3}{(2 (P_1 + P_2) - P_3 + 3 P_4)^2 (1 + \epsilon_E + \epsilon_I)} \nonumber\\
&  &-\frac{m_3^3 (1 + \epsilon_E + \epsilon_I)^3}{(2 P_3 - 2 P_4 - P_5 + P_6)^2 (1 + \epsilon_E + \epsilon_G + \epsilon_I)} \nonumber\\
&  &-\frac{J_2 R_{\jupiter}^2 m_1^7 (2 P_1 - P_2 - P_4 + P_5)}{2(-P_1 - P_2 - P_4 + P_5)^7 (1 + \epsilon_I)^3} \nonumber\\
&  &-\frac{J_2 R_{\jupiter}^2 m_2^7 (3 P_2 + 2 (P_1+ P_2) - P_3 + 3 P_4) (1 + \epsilon_I)^7}{(2 (P_1 + P_2) - P_3 + 3 P_4)^7 (1 + \epsilon_E + \epsilon_I)^3} \nonumber\\
&  &-\frac{J_2 R_{\jupiter}^2 m_3^7 (5 P_3 - 2 P_4 - P_5 + P_6) (1 + \epsilon_E + \epsilon_I)^7}{(2 P_3 - 2 P_4 - P_5 + P_6)^7 (1 + \epsilon_E + \epsilon_G + \epsilon_I)^3} \nonumber\\
&  &-\frac{m_1 m_2^2 \epsilon_E (1 + \epsilon_I)^2}{(2 (P_1 + P_2) - P_3 + 3 P_4)^2 (1 + \epsilon_E + \epsilon_I)} \times \nonumber\\
&  &\left(b_1 + \gamma_{11} \sqrt{\frac{2P_1}{-P_1 - P_2 - P_4 + P_5}} \cos q_1 +
   \gamma_{21} \sqrt{\frac{2P_2}{(2 (P_1 + P_2) - P_3 + 3 P_4)}} \cos q_2\right)      \nonumber\\
   &  &-\frac{m_2 m_3^2 \epsilon_G (1 +\epsilon_E + \epsilon_I)^2}{(2 P_3 - 2 P_4 - P_5 + P_6)^2 (1 + \epsilon_G+ \epsilon_E + \epsilon_I)} \times \nonumber\\
&  &\left(b_2 + \gamma_{12} \sqrt{\frac{2P_3}{2 P_3 - 2 P_4 - P_5 + P_6}} \cos q_3 +
   \gamma_{22} \sqrt{\frac{2P_2}{2 (P_1 + P_2) - P_3 + 3 P_4}} \cos (q_2-q_4)\right)   \nonumber\\
&  &-\frac{b_3 m_1 m_3^2 \epsilon_G (1 +\epsilon_E + \epsilon_I)^2}{(2 P_3 - 2 P_4 - P_5 + P_6)^2 (1 + \epsilon_G + \epsilon_E + \epsilon_I)} \ .   \nonumber\\
   \eeqa

In \equ{HBE} the following definitions and abbreviations have been introduced:
   \beqa{appdef}
   \epsilon_I &=& \frac{m_1}{m_0} \ , \label{MR1}\\
   \epsilon_E &=& \frac{m_2}{m_0} \ , \\
   \epsilon_G &=& \frac{m_3}{m_0} \ , \label{MR3}\\
   b_1 &=& {B_0}\left(\alpha_{12}\right) \ , \nonumber\\
   b_2 &=& {B_0}\left(\alpha_{23}\right) \ , \nonumber\\
   b_3 &=& {B_0}\left(\alpha_{13}\right) \ , \nonumber\\
   \gamma_{11} &=& \gamma_{1} \left(\alpha_{12}\right) \ , \nonumber\\
   \gamma_{12} &=& \gamma_{1} \left(\alpha_{23}\right) \ , \nonumber\\
   \gamma_{21} &=& \gamma_{2} \left(\alpha_{12}\right) \ , \nonumber\\
   \gamma_{22} &=& \gamma_{2} \left(\alpha_{23}\right)  \ .   \nonumber\\\nonumber
    \eeqa
    The mass ratios (\ref{MR1})--(\ref{MR3}) are reported in Table \ref{T1}, where the mass of Jupiter is taken as $m_0=1.898 \times 10^{27}$ kg.

\section{Second-order model}\label{app:B}

The extended model contains the following terms \cite{Ma}: the satellite self-interactions at 2nd order in the eccentricities
\beq{Hsat2}
H^{(2)}_{Sat}=-{{Gm_1m_2}\over a_2} R^{(12)}-{{Gm_2m_3}\over a_3} R^{(23)}-{{Gm_1m_3}\over a_3} R^{(13)} \ ,
\eeq
with
\beqa{}
R^{(12)}&=&
\gamma_0 \left(\alpha_{12}\right) (e_1^2+e_2^2) +
\gamma_3 \left(\alpha_{12}\right) e_1^2 \cos2(2\lambda_2-\lambda_1-\varpi_1)+
\gamma_4 \left(\alpha_{12}\right) e_2^2 \cos2(2\lambda_2-\lambda_1-\varpi_2) \ \nonumber\\
&&+e_1 e_2 \left[
\gamma_5  \left(\alpha_{12}\right) \cos(4\lambda_2-2\lambda_1-\varpi_2-\varpi_1) +
\gamma_6  \left(\alpha_{12}\right) \cos(\varpi_2-\varpi_1) \right]\ ,\nonumber\\
R^{(23)}&=&
\gamma_0 \left(\alpha_{23}\right) (e_2^2+e_3^2) +
\gamma_3 \left(\alpha_{23}\right) e_2^2 \cos2(\lambda_2-2\lambda_3+\varpi_2)+
\gamma_4 \left(\alpha_{23}\right) e_3^2 \cos2(\lambda_2-2\lambda_3+\varpi_3) \ \nonumber\\
&&+e_2 e_3 \left[\gamma_5  \left(\alpha_{23}\right) \cos(2\lambda_2-4\lambda_3+\varpi_2+\varpi_3) +
\gamma_6  \left(\alpha_{23}\right) \cos(\varpi_3-\varpi_2) \right]\ ,\nonumber\\
R^{(13)}&=&
\gamma_0 \left(\alpha_{13}\right) (e_1^2+e_3^2) +
\gamma_6 \left(\alpha_{13}\right) e_1 e_3 \cos(\varpi_3-\varpi_1)
\ , \nonumber\eeqa
where $\gamma_1$ and $\gamma_2$ are defined in \eqref{gammak} and
\beqano
\gamma_0&=&-{1\over 8}(2\alpha{d\over {d\alpha}}+\alpha^2{d^2\over {d\alpha^2}})b_{1/2}^{(0)}(\alpha)\ ,\qquad\qquad
\gamma_3=-{1\over 8}(44+14\alpha{d\over {d\alpha}}+\alpha^2{d^2\over {d\alpha^2}})b_{1/2}^{(4)}(\alpha)\ ,\nonumber\\
\gamma_4&=&-{1\over 8}(38+14\alpha{d\over {d\alpha}}+\alpha^2{d^2\over {d\alpha^2}})b_{1/2}^{(2)}(\alpha)\ ,\qquad
\gamma_5={1\over 4}(42+14\alpha{d\over {d\alpha}}+\alpha^2{d^2\over {d\alpha^2}})b_{1/2}^{(3)}(\alpha)\ ,\nonumber\\
\gamma_6&=&-{1\over 4}(2-2\alpha{d\over {d\alpha}}-\alpha^2{d^2\over {d\alpha^2}})b_{1/2}^{(1)}(\alpha) \ ,
\eeqano
the secular influence of Callisto and the Sun
\beqano
H_{p}&=&-{{Gm_1m_p}\over a_p}
\left\{{1\over 2}b_{1/2}^{(0)}
\left({a_1\over a_p} \right)-1+{1\over 8}{a_1\over a_p} b_{3/2}^{(1)}
\left({a_1\over a_p} \right)(e_1^2+e_p^2) \right\}\nonumber\\
&-&{{Gm_2m_p}\over a_p}
\left\{{1\over 2}b_{1/2}^{(0)}
\left({a_2\over a_p}\right)-1+{1\over 8}{a_2\over a_p} b_{3/2}^{(1)}
\left({a_2\over a_p}\right)(e_2^2+e_p^2)\right\}\nonumber\\
&-&{{Gm_3m_p}\over a_p}
\left\{{1\over 2}b_{1/2}^{(0)}
\left({a_3\over a_p}\right)-1+{1\over 8}{a_3\over a_p} b_{3/2}^{(1)}
\left({a_3\over a_p}\right)(e_3^2+e_p^2)\right\}\ ,
\eeqano
where the subindex $p$ denotes either Callisto or the Sun, while $m_p$, $a_p$, $e_p$ are, respectively,
the corresponding mass, semimajor axis, eccentricity
and the octupolar terms of Jupiter gravitational field
\beq{Hjup4}
H_{J_{4}}=
{{3 G m_0 J_4 R_{\jupiter}^4}\over {8}}\ \left[
{m_1\over a_1^4} \left(1+{5\over 2}e_1^2\right) +
{m_1+m_2\over a_2^4} \left(1+{5\over 2}e_2^2\right) +
{m_1+m_2+m_3\over a_3^4} \left(1+{5\over 2}e_3^2\right) \right] \ , \nonumber
\eeq
where $J_4 = -5.87 \times 10^{-4}$.
\bibliographystyle{alpha}
\bibliography{biblioCPPds}

\end{document}